\documentstyle[aps,twocolumn]{revtex}

\begin{document}
\title{Quantum Strategy Without Entanglement}
\author{Jiangfeng Du\thanks{%
Email: djf@ustc.edu.cn}, Xiaodong Xu\thanks{%
Email: xuxd@mail.ustc.edu.cn}, Hui Li\thanks{%
Email: lhuy@mail.ustc.edu.cn}, Mingjun Shi, Xianyi Zhou, Rongdian Han}
\address{{\small Laboratory of Quantum Communication and Quantum Computation,}\\
{\small Department of Modern Physics, University of Science and Technology}\\
{\small of China, Hefei, 230026, P.R.China}}
\date{Nov. 18, 2000}
\maketitle

\begin{abstract}
In this paper we quantize the Card Game. In the classical version of this
game, one player (Alice) can always win with propability 2/3. But when the
other player (Bob) is allowed to apply quantum strategy, the original unfair
game turns into a fair and zero-sum game. Further more, the procedure in
which Bob perform his quantum strategy does not include any ingredient of
entanglement.

PACS: \ 03.67.-a, 03.65.Bz, 02.50.Le

Key Words: \ Quantum Strategy, Entanglement, Quantum Game
\end{abstract}

\section{Introduction}

Game theory is the mathematical study of conflict situations, and it was, to
all intents and purposes, born in 1944 with the publication of a single book 
{\it Theory of Games and Economic Behavior} by J.Von Neumann and
O.Morgenstern. It now have a central place in economic theory, and it has
contributed important insights to all areas of social science.

Recently, Game Theory was extended into quantum world and the quantum
strategies were discussed\cite{1,2} and shown powerful\cite{3,4,5,6}.
L.Goldenberg, L.Vaidman and S.Wienser\cite{3} constructed a two-party
protocol for quantum gambling. In their protocol two remote parties were
allowed to play a gambling game, such that in a certain limit it becomes a
fair game. D.A.Meyer\cite{4} quantized the PQ Game, he found out that one
player could increase his expected payoff by implementing quantum strategy
against his classical opponent. J.Eisert, M.Wilkens and M. Lewenstein\cite{5}
presented the Prisoner's Dilemma. For the particular case they showed that
this game ceases to pose a dilemma if quantum strategies are allowed for.
S.C.Benjamin and P.M.Hayden\cite{6} quantized games with more than two
players. They demonstrate that such games can exhibit `coherent' equilibrium
strategies.

In this paper, we study an interesting two player Card Game. In the
classical game, one player (Alice) can always win with the probability $2/3$%
 But if the other player (Bob) performs quantum strategy, he can increase
his winning probability to $1/2$. Hence the unfair classical game becomes
fair in quantum world. In addition, we point out that this strategy does not
use entanglement, which is different from most of previous works.

In the following, we introduce the classical model of the Card Game at
first. Then we give the quantum scheme of the Card Game. Further, we show
that there is no entanglement in the quantum game.

\section{The Model of Card Game}

The classical model of card game is explained as following. Alice has three
cards. The first card has one circle in its both sides, the second has one
dot in its both sides and the third card has one circle in one side and one
dot in the other. At the first step, Alice put\ the three cards into a black
box. The cards are randomly placed in the box after Alice shakes it. Both
players cannot see what happens in the box. At the second step Bob take one
card from the box without flipping it. Both player can only see the upper
side of the card. Alice wins one coin if the pattern of the down side is the
same as that of the upper side and lose one coin when the patterns are
different. It is obvious that Alice has $2/3$ probability to win and Bob
only has $1/3$. So Bob is in a disadvantageous situation and the game is
unfair to him.

Any rational player will not play the game with Alice because the game is
unfair. In order to attract Bob to play with him, before the original second
step Alice allows Bob to has one chance to operate on the cards. That is Bob
has one step query on the box. In the classical world, Bob can only attain
one card information after the query. Because the card is in the box, so
what Bob knows is only one upper side pattern of the three cards. Except
this he know nothing about the three cards in the black box. So in the
classical field even having this one step query, Bob still will be in
disadvantageous state and the game is still unfair.

But when we investigate the game in the quantum field, the whole thing is
changed. We will see that the game turns into a fair zero-sum\cite{14} game
and both player are in equal situation.

\section{The Quantized Card Game}

In the first step, Alice puts the cards in the box and shake the box, that
is she prepares the initial state randomly. We describe the card state be $%
\left| 0\right\rangle $ if the pattern in the upper side is circle and $%
\left| 1\right\rangle $ if it is dot. So the upper sides of the three cards
in the box can be described as
\begin{equation}
\left| r\right\rangle =\left| r_{0}\right\rangle \left| r_{1}\right\rangle
\left| r_{2}\right\rangle
\end{equation}
where $r_{0},r_{1},r_{2}\in \left\{ 0,1\right\} $, which means $\left|
r_{0}\right\rangle $, $\left| r_{1}\right\rangle $ and $\left|
r_{2}\right\rangle $ are all eigenstate other than superposition of $\left|
0\right\rangle $ and $\left| 1\right\rangle $.

After the first step of the game, Alice give the black \ box to Bob. Because
Alice thinks in classical way, in her mind Bob cannot get information about
all upper side patterns of the three cards in the box. So she can still win
with higher probability. But what Bob use is quantum strategy. He replace
the classical one step query with one step quantum query. The following
shows how Bob query the box.

Bob has a quantum machine that applies an unitary operator $U$ on its three
input qubits and give three output qubits. This machine depends on the state 
$\left| r\right\rangle $ in the box that Alice gives Bob. The explicit
expression of $U$ and its relation with $\left| r\right\rangle $ is as
following.
\begin{equation}
U=U_{0}\otimes U_{1}\otimes U_{2}
\end{equation}
where
\begin{eqnarray}
U_{k} &=&\left\{ 
\begin{array}{l}
I_{2}=\left( 
\begin{array}{cc}
1 & 0 \\ 
0 & 1
\end{array}
\right) \text{ \ \ }if\text{ }r_{k}=0 \\ 
\sigma _{z}=\left( 
\begin{array}{cc}
1 & 0 \\ 
0 & -1
\end{array}
\right) \text{ }if\text{ }r_{k}=1
\end{array}
\right.  \nonumber \\
&=&\left( 
\begin{array}{cc}
1 & 0 \\ 
0 & e^{i\pi r_{k}}
\end{array}
\right)
\end{eqnarray}

The procession of query is shown in Figure 1. $H=\frac 1{\sqrt{2}}\left( 
\begin{array}{cc}
1 & 1 \\ 
1 & -1
\end{array}
\right) $ is Hadamard transformation and $U$ is the operator described above.

After the process, the output state is
\begin{eqnarray}
\left| \psi _{out}\right\rangle &=&\left( H\otimes H\otimes H\right) U\left(
H\otimes H\otimes H\right) \left| 0\right\rangle \left| 0\right\rangle
\left| 0\right\rangle  \nonumber \\
&=&(HU_{0}H\left| 0\right\rangle )\otimes (HU_{1}H\left| 0\right\rangle
)\otimes (HU_{2}H\left| 0\right\rangle )
\end{eqnarray}
Because
\begin{eqnarray}
HU_{k}H &=&\frac{1}{2}\left( 
\begin{array}{cc}
1 & 1 \\ 
1 & -1
\end{array}
\right) \left( 
\begin{array}{cc}
1 & 0 \\ 
0 & e^{i\pi r_{k}}
\end{array}
\right) \left( 
\begin{array}{cc}
1 & 1 \\ 
1 & -1
\end{array}
\right)  \nonumber \\
&=&\frac{1}{2}\left( 
\begin{array}{cc}
1+e^{i\pi r_{k}} & 1-e^{i\pi r_{k}} \\ 
1-e^{i\pi r_{k}} & 1+e^{i\pi r_{k}}
\end{array}
\right)
\end{eqnarray}
So
\begin{eqnarray}
HU_{k}H\left| 0\right\rangle &=&\frac{1+e^{i\pi r_{k}}}{2}\left|
0\right\rangle +\frac{1-e^{i\pi r_{k}}}{2}\left| 1\right\rangle  \nonumber \\
&=&\left\{ 
\begin{array}{c}
\left| 0\right\rangle \text{ \ \ }if\text{ }r_{k}=0 \\ 
\left| 1\right\rangle \text{ \ \ }if\text{ }r_{k}=1
\end{array}
\right.  \nonumber \\
&=&\left| r_{k}\right\rangle
\end{eqnarray}

From above, it is obvious to see that Bob can obtain the complete
information about the upper patterns of all the three cards through only one
query. There are only two possible kinds of output states in the black box,
which is $\left| 0\right\rangle \left| 0\right\rangle \left| 1\right\rangle $
or $\left| 1\right\rangle \left| 1\right\rangle \left| 0\right\rangle $,
that is two circles and one dot in the upper side or two dots and one circle
(here three cards have no sequence between each other, for example, $\left|
0\right\rangle \left| 0\right\rangle \left| 1\right\rangle $ is the same as $%
\left| 0\right\rangle \left| 1\right\rangle \left| 0\right\rangle $ and $%
\left| 1\right\rangle \left| 0\right\rangle \left| 0\right\rangle $). For
the convenience of explanation, we assume that the states of the cards after
first step is two circles and one dot ($\left| 0\right\rangle \left|
0\right\rangle \left| 1\right\rangle $). After one step query, Bob knows the
complete information about the upper patterns, but he has no individual
information about which upper pattern corresponding to which card. Then he
takes one card out of the box and see what pattern is in the upper side. If
he finds out that he is in the disadvantage situation, the upper pattern of
the card is dot ($\left| 1\right\rangle $), he refuses to play with Alice in
this turn because he know the down side pattern is dot definitely. Otherwise
if the upper side pattern is circle ($\left| 0\right\rangle $), then he
knows that the down side pattern is either circle $\left| 0\right\rangle $
or dot $\left| 1\right\rangle $. So he continue this turn because he has the
probability $\frac 12$ to win. Bob will continue the game because he has
probability $\frac 12$ to win. Hence the game becomes fair and is also
zero-sum.

\section{No Entanglement Applied}

Entanglement is regarded as the crucial ingredient in quantum information
and quantum computation\cite{7,8,9,10}. And in quantum games of previous
works, most quantum strategies exceed classical ones because of the power of
entanglement. In J.Eisert, M.Wilkens and M. Lewenstein\cite{5} and
S.C.Benjamin and P.M.Hayden\cite{6} works, they used the $J$ gate to
entangle the initial state. And one of the reason why their quantum
strategies are better than classical strategies is that the initial state is
max-entangled.

Recently, some works have shown that entanglement is not necessary for
information processing and algorithm\cite{11,12,13,15}. This is also true in
quantum game and quantum strategy. In D.A.Meyer's work of PQ-Game\cite{4,16}%
, there is no entanglement in the strategies and the quantum game is still
exceed its classical version. In this paper, the quantum strategy applied by
Bob includes no entanglement and is still better than classical strategy.

The initial state input the quantum machine is $\left| 0\right\rangle \left|
0\right\rangle \left| 0\right\rangle $, which is obviously separable. After
the Hadamard transformation, the state is
\[
\frac 1{\sqrt{2^3}}\left( \left| 0\right\rangle +\left| 1\right\rangle
\right) \otimes \left( \left| 0\right\rangle +\left| 1\right\rangle \right)
\otimes \left( \left| 0\right\rangle +\left| 1\right\rangle \right) 
\]
Performed by $U$, the state becomes
\[
\frac 1{\sqrt{2^3}}\left( \left| 0\right\rangle +e^{i\pi r_0}\left|
1\right\rangle \right) \otimes \left( \left| 0\right\rangle +e^{i\pi
r_1}\left| 1\right\rangle \right) \otimes \left( \left| 0\right\rangle
+e^{i\pi r_2}\left| 1\right\rangle \right) 
\]
And the states after the second Hadamard transformation is the output state
\[
\left| r_0\right\rangle \left| r_1\right\rangle \left| r_2\right\rangle 
\]

In the whole procedure, the state is tensor products of the states of the
individual qubits, so it is unentangled.. And because the operators ($H$ and 
$U$) are also tensor product of the individual local operators on these
qubits, so it is obviously that in this quantum game there is no
entanglement applied.

We can see that, from previous works, entanglement is important for static
games (such as Prisoner'sss Dilemma\cite{5} and Battle of The Sexes Game\cite
{17,18}), but may be not necessary in dynamic games (such as the PQ-Game\cite
{16} and the Card Game in this paper). Why entanglement is the crucial
ingredient in some quantum games but is not in others? In static games, each
player can only control his qubit and his operation is local. So in
classical world, the operation of one player cannot have  influence on
others in the operational process. But in quantum field, we think that
through entanglement the strategy-changing of one player could influent not
only himself but also his opponents. In dynamic games, players can control
all qubits at any time step. So just like quantum algorithms\cite{15}, in
dynamic games players can use quantum strategies without entanglement to
solve problem, even entangled quantum strategies could be redescribed with
other quantum strategies without entanglement. 

\section{Conclusion}

In this paper, we generalize the classical card game into quantum world. We
show that if Bob is given quantum strategy --- one step quantum query ---
against his classical opponent Alice, she cannot always win with high
probability. Both players are in equal situation and the game is a fair
zero-sum game.

Further, in this paper, the quantum Card Game includes no entanglement. The
quantum-over-classical strategy is achieved using only interference. Like in
quantum information and algorithm, entanglement may be not necessary in
quantum strategy. Quantum strategy could be still powerful without
entanglement.

{\bf Acknowledgment:} This project is supported by the National Nature
Science Foundation of China(No.10075041 and No.10075044) and the Science
Foundation for Young Scientists of USTC.

\section{Figure Caption}

The process through which Bob attain the information about the state of
cards. Where $H$ is Hadamard transformation and $U$ is the operator Bob
applied.

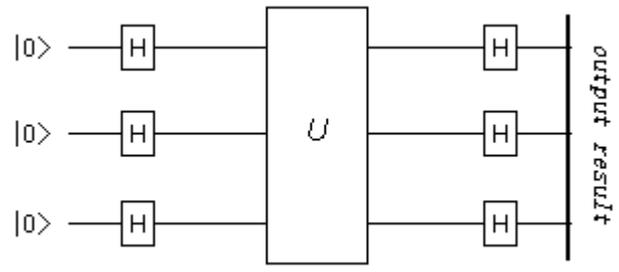
\end{document}